\documentclass[conference]{IEEEtran}

\IEEEoverridecommandlockouts

\usepackage[english]{babel}
\usepackage{glossaries}
\usepackage{amsmath}
\usepackage{amsfonts}
\usepackage[dvipsnames]{xcolor}
\usepackage{hyperref}
\usepackage{cleveref}
\usepackage{url}
\usepackage{pgfplots}
\usepackage{pgfplotstable}
\usepackage{tikz}
\usepackage[inkscapelatex=false]{svg}
\usepackage{cite}
\usepackage[activate={true,nocompatibility},tracking=true,final]{microtype}
\usepackage[per-mode=repeated-symbol,retain-explicit-plus=true]{siunitx}
\usepackage{xspace}
\usepackage{booktabs}
\usepackage{multirow}
\usepackage{subcaption}
\usepackage{pifont}
\usepackage{tablefootnote}
\usepackage{circledsteps}

\newcommand{\cmark}{{\ding{52}}}
\newcommand{\xmark}{{\ding{56}}}

\newif\ifblind
\blindfalse

\newacronym[longplural={Scratchpad Memories}]{SPM}{SPM}{Scratchpad Memory}
\newacronym[longplural={Standard Cell Memories}]{SCM}{SCM}{Standard Cell Memory}
\newacronym[longplural={Static Random-Access Memories}]{SRAM}{SRAM}{Static Random-Access Memory}
\newacronym{1RW}{1RW}{single read/write port}
\newacronym{3DIC}{3D-IC}{three-dimensional integrated circuit}
\newacronym{3R1W}{3R1W}{three read ports and one write port}
\newacronym{ACE}{ACE}{AXI Coherent Extensions}
\newacronym{AI}{AI}{Artificial Intelligence}
\newacronym{AMBA}{AMBA}{Advanced Microcontroller Bus Architecture}
\newacronym{APB}{APB}{Advanced Peripheral Bus}
\newacronym{API}{API}{Application Programming Interface}
\newacronym{ASIC}{ASIC}{Application-Specific Integrated Circuit}
\newacronym{AVX}{AVX}{Advanced Vector Extension}
\newacronym{AXI}{AXI4}{Advanced eXtensible Interface}
\newacronym{BEOL}{BEOL}{back end of the line}
\newacronym{BLAS}{BLAS}{Basic Linear Algebra Subprograms}
\newacronym{C4}{C4}{controlled collapse chip connection}
\newacronym[longplural={Core Complexes}]{CC}{CC}{Core Complex}
\newacronym{CHI}{CHI}{Coherent Hub Interface}
\newacronym{CMG}{CMG}{Core Memory Group}
\newacronym{CMOS}{CMOS}{Complementary Metal-Oxide-Semi\-con\-ductor}
\newacronym{CNN}{CNN}{Convolutional Neural Network}
\newacronym{CPU}{CPU}{Central Processing Unit}
\newacronym{CSR}{CSR}{Control and State Register}
\newacronym{CTS}{CTS}{Clock Tree Synthesis}
\newacronym{DCT}{DCT}{discrete cosine transform}
\newacronym{DDR}{DDR}{double data rate}
\newacronym{DDol}{D\$}{Data Cache}
\newacronym{DLP}{DLP}{Data Level Parallelism}
\newacronym{DMA}{DMA}{Direct Memory Access}
\newacronym{DRAM}{DRAM}{Dynamic Random-Access Memory}
\newacronym{DRV}{DRV}{design rule violation}
\newacronym{DSP}{DSP}{Digital Signal Processing}
\newacronym{DUT}{DUT}{Device Under Test}
\newacronym{ECL}{ECL}{Emitter-Coupled Logic}
\newacronym{EDP}{EDP}{energy-delay product}
\newacronym{F2F}{F2F}{face-to-face}
\newacronym{FBB}{FBB}{Forward Body-Biasing}
\newacronym{FDSOI}{FD-SOI}{Fully Depleted Silicon on Insulator}
\newacronym{FEOL}{FEOL}{front end of the line}
\newacronym{FF}{FF}{Flip-Flops}
\newacronym{FMA}{FMA}{Fused Multiply-Add}
\newacronym{FO4}{FO4}{Fanout-of-4 Inverter}
\newacronym{FPGA}{FPGA}{Field-Pro\-gram\-ma\-ble Gate Array}
\newacronym{FPRF}{FPR}{Floating-Point Register File}
\newacronym{FU}{FU}{Functional Unit}
\newacronym{FPU}{FPU}{Floating Point Unit}
\newacronym{GPGPU}{GPGPU}{General-Purpose \acrlong{GPU}}
\newacronym{GPRF}{GPR}{General-Purpose Register File}
\newacronym{GPU}{GPU}{Graphics Processing Unit}
\newacronym{HBM}{HBM}{High Bandwidth Memory}
\newacronym{HDL}{HDL}{Hardware Description Language}
\newacronym{HERO}{HERO}{Heterogeneous Embedded Research Platform}
\newacronym{HPC}{HPC}{High-Performance Computing}
\newacronym{IC}{IC}{integrated circuit}
\newacronym{ICG}{ICG}{Integrated Clock Gating}
\newacronym{IDol}{I\$}{Instruction Cache}
\newacronym{ILP}{ILP}{Instruction Level Parallelism}
\newacronym{IOT}{IoT}{Internet of Things}
\newacronym{IPC}{IPC}{Instructions Per Cycle}
\newacronym{IPU}{IPU}{Integer Processing Unit}
\newacronym{ISA}{ISA}{Instruction Set Architecture}
\newacronym{LMUL}{LMUL}{Vector Length Multiplier}
\newacronym{LSU}{LSU}{Load/Store Unit}
\newacronym{LVT}{LVT}{low voltage threshold}
\newacronym{MACU}{MACU}{Multiply-Accumulate Unit}
\newacronym{MIMD}{MIMD}{Multiple Instruction, Multiple Data}
\newacronym{ML}{ML}{Machine Learning}
\newacronym{MMU}{MMU}{Memory Management Unit}
\newacronym{MUL}{MUL}{multiplier}
\newacronym{MVE}{MVE}{M-Profile Vector Extension}
\newacronym{MVL}{MVL}{maximum vector length}
\newacronym{NI}{NI}{Network Interface}
\newacronym[longplural={Networks-on-Chip}]{NOC}{NoC}{Network-on-Chip}
\newacronym{NUMA}{NUMA}{non-uniform memory access}
\newacronym{PCIe}{PCIe}{Peripheral Component Interconnect Express}
\newacronym{PC}{PC}{Program Counter}
\newacronym{PDP}{PDP}{power-delay product}
\newacronym{PE}{PE}{Processing Element}
\newacronym{PL}{PL}{Programmable Logic}
\newacronym{PMCA}{PMCA}{Programmable Manycore Accelerator}
\newacronym{PPA}{PPA}{power, performance, and area}
\newacronym{PSL}{PSL}{Power Service Layer}
\newacronym{PTE}{PTE}{page-table entry}
\newacronym{PTW}{PTW}{page-table walker}
\newacronym{PULP}{PULP}{Parallel Ultra Low Power}
\newacronym{RAM}{RAM}{Random-Access Memory}
\newacronym{RAW}{RAW}{read-after-write}
\newacronym{RBB}{RBB}{Reverse Body-Biasing}
\newacronym{ROB}{ROB}{Reorder Buffer}
\newacronym{RTL}{RTL}{Register Transfer Level}
\newacronym{RVT}{RVT}{Regular Voltage Threshold}
\newacronym{RVV}{RVV}{RISC-V Vector Extension}
\newacronym{RoCC}{RoCC}{Rocket Custom Coprocessor Interface}
\newacronym{SDRAM}{SDRAM}{synchronous dynamic random-access memory}
\newacronym{SIMD}{SIMD}{Single Instruction, Multiple Data}
\newacronym{SIMT}{SIMT}{Single Instruction, Multiple Thread}
\newacronym{SLDU}{SLDU}{Slide Unit}
\newacronym{SLVT}{SLVT}{super-low voltage threshold}
\newacronym{SM}{SM}{Streaming Multiprocessor}
\newacronym{SOC}{SoC}{System-on-Chip}
\newacronym{SSE}{SSE}{Streaming SIMD Extension}
\newacronym{SSR}{SSR}{Stream Semantic Register}
\newacronym{STA}{STA}{Static Timing Analysis}
\newacronym{STCO}{STCO}{System-Technology Co-Optimization}
\newacronym{SVE}{SVE}{Scalable Vector Extension}
\newacronym{TLP}{TLP}{Thread Level Parallelism}
\newacronym{TSV}{TSV}{through-silicon via}
\newacronym{TxnID}{TxnID}{Transaction ID}
\newacronym{VAC}{VAC}{Vector Access}
\newacronym{VAU}{VAU}{Vector Arithmetic Unit}
\newacronym{VCONV}{VCONV}{Vector Conversion}
\newacronym{VC}{VC}{virtual channel}
\newacronym{VEX}{VEX}{Vector Execute}
\newacronym{VFU}{VFU}{vector functional unit}
\newacronym{VID}{VID}{Vector Instruction Decode}
\newacronym{VIS}{VISSUE}{Vector Instruction Issue}
\newacronym{VLA}{VLA}{Vector-Length Agnostic}
\newacronym{VLSI}{VLSI}{Very Large-Scale Integration}
\newacronym{VLIW}{VLIW}{Very Long Instruction Word}
\newacronym{VLOOP}{VLOOP}{Vector Loop}
\newacronym{VLR}{VLR}{vector length register}
\newacronym{VLSU}{VLSU}{Vector Load/Store Unit}
\newacronym{VNB}{VNB}{Von Neumann Bottleneck}
\newacronym{VPU}{VPU}{Vector Processing Unit}
\newacronym{VRF}{VRF}{Vector Register File}
\newacronym{VSLDU}{VSLDU}{Vector Slide Unit}
\newacronym{VT}{VT}{vector thread}
\newacronym{W2W}{W2W}{wafer-to-wafer}
\newacronym{WAR}{WAR}{write-after-read}
\newacronym{WAW}{WAW}{write-after-write}

\definecolor{PULPRed}{HTML}{A8322C}
\definecolor{PULPBlue}{HTML}{1269B0}
\definecolor{PULPGreen}{HTML}{168638}
\definecolor{PULPOrange}{HTML}{F29545}
\definecolor{PULPPurple}{HTML}{910569}
\definecolor{PULPOlive}{HTML}{48592C}
\definecolor{PULPMarine}{HTML}{007996}
\definecolor{PULPGray}{HTML}{ABABAB}
\definecolor{Red}{HTML}{FF0000}

\colorlet{color1}{PULPBlue}
\colorlet{color2}{PULPRed}
\colorlet{color3}{PULPGreen}
\colorlet{color4}{PULPOrange}
\colorlet{color5}{PULPPurple}
\colorlet{color6}{PULPOlive}
\colorlet{color7}{PULPMarine}

\colorlet{colorCore}{PULPRed}
\colorlet{colorMemory}{PULPBlue}
\colorlet{colorInterconnect}{PULPGreen}
\colorlet{colorAccelerator}{PULPOrange}
\colorlet{colorPeripheral}{PULPPurple}
\colorlet{colorAlert}{Red}

\DeclareSIUnit\bank{bank}
\DeclareSIUnit\cycle{cycle}
\DeclareSIQualifier\double{DP}
\DeclareSIQualifier\single{SP}
\DeclareSIUnit\flop{FLOP}
\DeclareSIUnit\flops{FLOPS}
\DeclareSIUnit\gate{GE}
\DeclareSIUnit\op{OP}
\DeclareSIUnit\ops{OPS}
\DeclareSIUnit\bitpersecond{bps}
\DeclareSIUnit\request{request}
\DeclareSIUnit\core{core}
\DeclareSIUnit\pin{pin}
\DeclareSIUnit\hop{hop}
\DeclareSIUnit\byte{B}
\DeclareSIUnit\bit{bit}
\DeclareSIUnit\percent{\%}

\usetikzlibrary{scopes, calc, shapes, arrows, positioning, patterns}
\tikzset{>=latex}

\pgfplotsset{compat=1.17}
\pgfplotsset{width=\linewidth, height=7cm}
\pgfplotsset{every x tick label/.append style={font=\small}}
\pgfplotsset{every y tick label/.append style={font=\small}}

\SetTracking{encoding={*}, shape=sc}{0}

\begin{document}

\title{FlooNoC: A Multi-Tbps Wide NoC for Heterogeneous AXI4 Traffic
}

\author{
\ifblind
    Blind for review
\else
    \IEEEauthorblockN{Tim Fischer\IEEEauthorrefmark{1}, Michael Rogenmoser\IEEEauthorrefmark{1}, Matheus Cavalcante\IEEEauthorrefmark{1}, Frank K. Gürkaynak\IEEEauthorrefmark{1} Luca Benini\IEEEauthorrefmark{1}\IEEEauthorrefmark{2}}
    \IEEEauthorblockA{\IEEEauthorrefmark{1}Integrated Systems Laboratory (IIS), ETH Zurich
    \IEEEauthorblockA{\IEEEauthorrefmark{2}Department of Electrical, Electronic and Information Engineering (DEI), University of Bologna, Italy
    \\\{fischeti, michaero, matheus, kgf, lbenini\}@iis.ee.ethz.ch}}
\fi
}
\maketitle

\begin{abstract}
Meeting the staggering bandwidth requirements of today’s
applications challenges the traditional narrow and serialized NoCs,
which hit hard bounds on the maximum
operating frequency. This paper proposes FlooNoC, an open-source,
low-latency, fully AXI4-compatible NoC with wide physical channels for latency-tolerant high-bandwidth non-blocking transactions and decoupled latency-critical short  messages. We demonstrate the
feasibility of wide channels by integrating a 5$\times$5 router and links
within a 9-core compute cluster in 12\,nm FinFet technology.
Our NoC achieves a bandwidth of 629\,Gbps per link while
running at only \SI{1.23}{\giga\hertz} (at \SI{0.19}{\pico\joule\per\byte\per\hop}), with
just 10\% area overhead post layout.
\end{abstract}


\bstctlcite{IEEE:BSTcontrol}

\section{Introduction}
The ever-growing demand for computing resources is accompanied by a corresponding increase in the number of cores and \glspl{PE} instantiated in modern \glspl{SOC}. As a result, \glspl{NOC} must keep up with the consequent increase in communication bandwidth to avoid becoming a bottleneck for the overall performance and energy. Worse yet, most of today's application traffic is memory-bound and suffers from high latency due to off-chip transfers via memory controllers. To alleviate this problem, \glspl{NOC} must support numerous outstanding transactions and burst-based data transfers. This way, \glspl{PE} remain occupied, and the available memory bandwidth is fully utilized. 

The leading standard for on-chip communication supporting bursts and multiple outstanding transactions for non-coherent data transfers is \gls{AXI}. However, its strict ordering requirement on transactions with the same ID makes it very challenging for \gls{AXI} to be used as a link-level protocol for a \gls{NOC}. The logic required to track outstanding transactions and adhere to the ordering constraints increases exponentially in complexity~\cite{kurth_open-source_2022} with the network diameter. Therefore, the protocol overhead limits the scalability of multi-hop interconnects with \gls{AXI}-compliant links.

Decoupling the \gls{NOC}'s link-level protocol from the \gls{AXI}-compliant initiator and target \gls{NI}~\cite{benini_networks_2002, jerger_-chip_2017} helps overcome \gls{AXI}'s intrinsic scalability issue and enables flexible and scalable interconnect architectures for many-core systems. In particular, a custom \gls{NOC} link-level protocol allows the streamlining of the router architecture. For example, packets can be forwarded based solely on their destination address, while the \glspl{NI} manage outstanding transactions and order network responses. However, despite the proliferation of open-source \glspl{NOC}~\cite{esp_noc_soc, celerity, openpiton, zhao_constellation_2022}, \gls{AXI}-compliant \glspl{NI} with support for multiple outstanding transactions and bursts are currently only available as closed-source industry solutions~\cite{arm_noc, flexnoc}.

In modern high-bandwidth interconnects, end-to-end messages are typically large, in the order of a few \SI{}{\kilo\byte}. The traditional approach of sending messages through a \gls{NOC} has been to serialize them over a narrow link~\cite{jerger_-chip_2017}. 
Therefore, \glspl{NOC} meet their high throughput requirements by increasing their operating frequency. However, this approach becomes untenable for modern systems where the bandwidth injected by the endpoints is very high. For instance, serializing a 512-bit \gls{AXI} channel running at a moderate \SI{1}{\giga\hertz} onto a narrow 32-bit \gls{NOC} link without bandwidth reduction would require a link frequency of \SI{16}{\giga\hertz}, which is unattainable due to physical implementation constraints. Therefore, the classical \textit{narrow serialized link} approach must be revised in modern \glspl{NOC}, and network links are bound to be wide because of maximum achievable clock frequency limitations.

Typical \glspl{NOC} share access to physical links through virtual channels. However, the need for such resource sharing is becoming increasingly untenable in modern \gls{VLSI} technologies~\cite{dally_keynote}, which have plenty of routing resources to support multiple wide physical links. Even though virtual channels improve the network's physical routability thanks to their reduced wire count, they also reduce the effective bandwidth and add complexity to the routers since they require additional buffering and multiplexing of virtual channels. Such factors suggest that physical links should be favored over virtual channels in modern technologies~\cite{dally_keynote}. Still, \glspl{NOC} with wide physical links must be designed with a strong awareness of physical implementation effects.

As an additional challenge, modern systems, especially heterogeneous ones, present extremely diverse traffic characteristics regarding bandwidth requirements and latency sensitivity. As a result, it is particularly challenging for interconnects to answer adequately to the equally diverse requirements of such traffic patterns. For instance, burst-based traffic generated by a \gls{DMA} engine is more latency tolerant and benefits from wide buses. 
On the other hand, synchronization and communication messages generated by controllers are typically composed of short and highly latency-sensitive messages. 
Unfortunately, there is no one-size-fits-all solution to handle such heterogeneous traffic patterns. Burst-based traffic has the potential to starve latency-sensitive communication due to its high bus utilization, while prioritizing small latency-sensitive messages on the network will degrade its peak throughput, which is needed for bulk data transfers.

To address these issues, we propose a new \gls{NOC} design based on the following principles: \textbf{1) Full \gls{AXI} support with multiple outstanding burst transactions}, with low-complexity routers and a decoupled link-level protocol to allow scalability, which is essential to tolerate high-latency off-chip accesses; \textbf{2) Wide physical channels} to satisfy the need for high bandwidth at the network endpoints, without being bound by the operating frequency; \textbf{3) Decoupled links and networks} 
for wide, high-bandwidth, and narrow, latency-sensitive traffic.
The contributions of this paper are the following:

\begin{itemize}
    \item We present the first (to the best of our knowledge) open-source\ifblind\else\footnote{https://github.com/pulp-platform/FlooNoC}\fi \gls{NOC} with fully \gls{AXI}-compliant initiator and target interfaces that efficiently handle the ordering requirements imposed by \gls{AXI} at the endpoint rather than in the routers while achieving full bandwidth utilization.
    \item We propose a method to separate and map diverse kinds of traffic on multiple links with different widths and demonstrate its efficacy in handling both high-bandwidth burst-based and latency-intolerant traffic.
    \item We integrate the proposed \gls{NOC} into an L1-shared compute cluster with 8 RISC-V cores~\cite{snitch}, forming a compute tile that can be scaled to large, complex, many-core \glspl{SOC}. We demonstrate the physical implementation of the merged compute and \gls{NOC} in a \SI{12}{\nano\meter} \gls{VLSI} technology node, achieving high energy efficiency of \SI{0.19}{\pico\joule\per\byte\per\hop} at \SI{1.23}{\giga\hertz}, where the area cost of our \gls{NOC} is \SI{450}{\kilo\gate}, contributing only 10\% to the area of the tile.
\end{itemize}

\section{Background}

While being the de-facto standard protocol for non-coherent on-chip communication, \gls{AXI} has some limitations when used as a link-level protocol.

\subsection{Advanced eXtensible Interface (AXI4)}
\gls{AXI} is an industry-standard protocol widely used for high-bandwidth on-chip communication. \Gls{AXI} defines separate channels for read and write requests (AR, AW, W) and response channels (R, B). The \gls{AXI} protocol supports many outstanding transactions, which are critical in systems with high-latency communication, such as I/O and off-chip memory accesses. Outstanding transactions allow the initiator to hide the memory latency, leading to higher system throughput. 

The \gls{AXI} protocol imposes strict ordering requirements based on the transaction ID. The ID field is an identifier for transactions, and its width depends on the number of initiators in the system and their characteristics. The \gls{AXI} protocol specification~\cite{axi_spec} requires that transactions with the same ID are always returned in order. Furthermore, as the ID is also used for routing a response, the protocol requires that the ID width must increase at each interconnection (hop) to retain the uniqueness of the transactions and their IDs. This presents a major challenge when scaling up \gls{AXI} to large and complex systems with many hops between initiator and targets, as state information is required for each ID for ordering while the number of possible IDs is growing. Tracking outstanding transactions based on the ID has exponential complexity~\cite{kurth_open-source_2022}, ultimately preventing scalability. There are measures to mitigate this issue, such as ID remappers~\cite{kurth_open-source_2022}, that decrease the number of IDs. However, these impose significant overhead in latency and area and greatly complicate tracing and verification. 



\section{NoC Architecture}
Due to these limitations of \gls{AXI}, we decouple the \gls{NOC} link-level protocol with an \gls{NI} to enable the scalability of the \gls{NOC}.

\subsection{Network Interface}

\begin{figure}[htbp]
    \centering
    \includegraphics[width=\columnwidth]{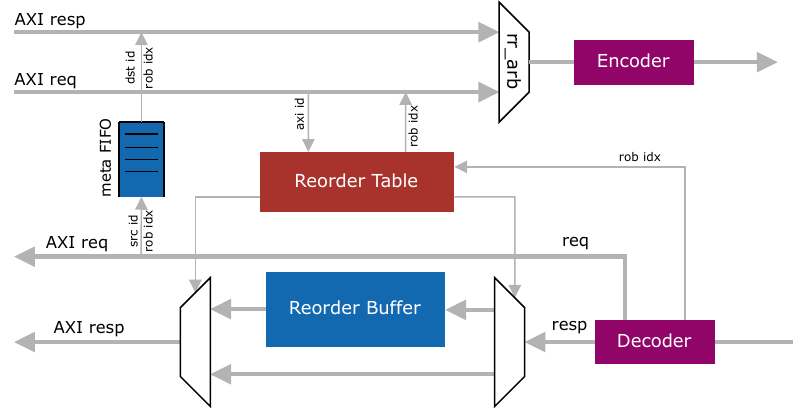}
    \caption{\gls{AXI} Network Interface with ROB to support the reordering requirements of \gls{AXI}.}
    \label{fig:network-interface}
\end{figure}

\gls{AXI} requires ordering of transactions with the same ID, often implemented using additional hardware in every switch to track outstanding transactions of every ID~\cite{kurth_open-source_2022}. To avoid this complexity, we handle reordering at the endpoint (\gls{NI}), allowing us to use simpler \gls{NOC} routers that do not need to guarantee in-order transactions. The mechanism for ordering responses is implemented in the \gls{NI} and is shown in \Cref{fig:network-interface}.

A reorder table keeps track of outstanding transactions for every ID and accepts new transactions using end-to-end flow control, meaning new \gls{AXI} requests are only injected into the \gls{NOC} if there is enough space in the \gls{ROB} to store the response. 

To mitigate large stalls that result from limited \gls{ROB} storage, we implement two optimizations 1) For a given ID, the first response of a stream of multiple transactions never needs to be reordered and can be directly forwarded to the \gls{AXI} interface; 2) Assuming deterministic routing in the network, the responses of requests to the same destination will arrive in the same order as the requests were issued. Hence there is no need to reorder them. In order to know if a response arrives in order, a unique identifier is sent with each request and returned by the responses. Based on this identifier, the reorder table can determine if the response should be buffered or directly forwarded to the \gls{AXI} interface. The unique identifier is the index into the \gls{ROB}, which was allocated when the request was granted. The \gls{ROB} allocation is dynamic and supports bursts of arbitrary lengths. The reorder table, which is used for the \gls{ROB} management, consists of a FIFO for each \gls{AXI} ID that can hold a configurable number of indexes into the \gls{ROB} (the depth corresponds to the number of outstanding transactions for each ID). Once a new outgoing \gls{AXI} request arrives, the next available \gls{ROB} space is checked, which can hold the size of the corresponding response. The index of the allocated \gls{ROB} space is pushed into the corresponding FIFO, and the space is reserved until the response has arrived. Each \gls{AXI} ID is managed separately. Hence responses from different \gls{AXI} can still be handled out of order.

To route the response of an incoming request back to the initiator, the source ID of the request is stored in the \textit{meta FIFO}, together with the information required for ordering the response. The order of all incoming non-atomic responses is preserved by serializing them with an identical \gls{AXI} ID. Atomic requests, on the other hand, require unique IDs and are stored in separate \textit{meta buffers} not depicted in \Cref{fig:network-interface}.

The \gls{AXI}-\gls{NI} also supports outstanding transactions, atomic accesses, and bursts. During a burst, each data beat is seamlessly sent as a flit in a single cycle, given no backpressure from the \gls{NOC} occurs. The \gls{NI} also supports multiple static routing algorithms like XY-Routing or table-based routing using the destination's ID. The destination and the source ID are encoded in the header to route back responses to the initiator.

\subsection{Narrow-Wide Links}

\begin{figure}[htbp]
    \centering
    \includegraphics[width=\columnwidth]{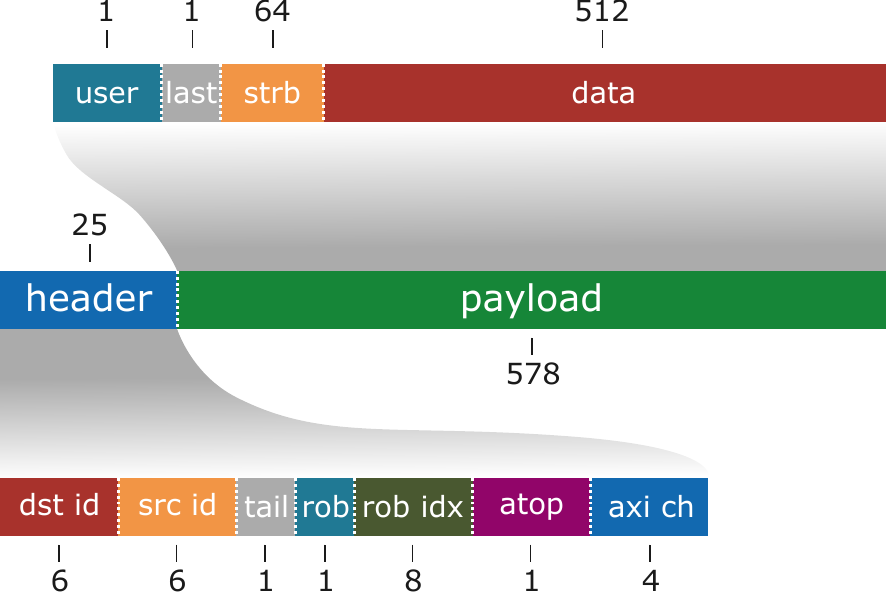}
    \caption{Example of a single flit, consisting of header information and an AXI W beat payload of 512-bit.}
    \label{fig:packet}
\end{figure}

Traditional \glspl{NOC} serialize a packet onto a narrow link and decode the packet's start and end with header and tail flits, with additional information, such as the destination in the header flit. Header and tail flits are costly when using wide physical links, as each packet is sent in a single cycle, limiting the effective bandwidth of single-packet transfers to 33\%. Instead, we use parallel lines for the header bits, such as routing, ordering, and payload type information, as depicted in \Cref{fig:packet}. The payload itself can be arbitrary, in our case it consists of \gls{AXI} requests and responses transmitted between initiator and target. 

The traffic through an \gls{SOC} interconnect can be very heterogeneous due to the many types of initiators involved. Programmable \glspl{DMA} are commonly used with very wide buses and burst-based data transfers to satisfy the need for bandwidth at the compute elements. Compute cores, on the other hand, typically only generate single-word transfers for tasks like synchronization and configuration. We use separate physical links for narrow, latency-sensitive, and wide, high-bandwidth traffic to accommodate various types of communication in our network. The sizes of both narrow (64-bit) and wide (512-bit) links are dimensioned such that all packets sent over a link fit into one flit which can be transmitted in a single cycle. This allows us to match the frequency and bandwidth of the endpoint \gls{AXI} bus and \gls{NOC} links. We implement three physical links in each direction, as listed in \Cref{tab:channels}. 

\begin{table}[htbp]
    \centering
    \caption{Description and dimensions of physical links. Mapping of AXI requests and responses of \textsc{DataWidth} = 64/512-bit, \textsc{AddrWidth} = 48-bit.}
    \begin{tabular}{@{}lccc@{}}
    \toprule
    & & \multicolumn{2}{c}{\textbf{Mapping \& Primary Payload}} \\
    \cmidrule(l){3-4}
    \textbf{Phys. link} & \textbf{Size} & \textbf{AXI Narrow} & \textbf{AXI Wide} \\
    \midrule
    \multirow{2}{*}{$\mathtt{narrow\_req}$} & \multirow{2}{*}{\SI{119}{\bit}} & AR/AW: 48-bit addr & \multirow{2}{*}{AR/AW: 48-bit addr} \\
    & & W: 64-bit data \\
    \midrule
    \multirow{2}{*}{$\mathtt{narrow\_rsp}$} & \multirow{2}{*}{\SI{103}{\bit}} & R: 64-bit data & \multirow{2}{*}{B: 2-bit resp}\\
    & & B: 2-bit resp &  \\
    \midrule
    \multirow{2}{*}{$\mathtt{wide}$} & \multirow{2}{*}{\SI{603}{\bit}} & \multirow{2}{*}{-} & W: 512-bit data \\
    & & & R: 512-bit data \\
    \bottomrule
    \\
    \end{tabular}
    \label{tab:channels}
\end{table}

\gls{AXI} requests and responses are always sent over different physical links to prevent message-level deadlocks. Further, $\mathtt{narrow\_req}$ and $\mathtt{narrow\_rsp}$ are primarily used to handle latency-sensitive requests and responses from the compute cores. Additionally, read and write requests and write responses of the wide \gls{AXI} bus are also mapped onto the narrow links since these messages would only utilize a fraction of the wide link. Mapping them to the narrow link frees up the wide link for high bandwidth traffic such as read and write bursts.

\subsection{Router}
Our \gls{NOC} benefits from wide physical links and reduced operating frequency compared to traditional narrow links, resulting in significant microarchitectural advantages for the router design. We use simple low-area and low-complexity routers without internal pipelining and virtual channels. Instead, we use multilink routers, which contain different routers for each of the three physical links, thus separating the networks completely from one another. Furthermore, the routers do not enforce ordering and exhibit much better scalability than interconnects based on \gls{AXI} matrices. Our routers are configurable to an arbitrary number of input and output ports, and we designed the routers to use wormhole routing with a valid-ready handshake for control flow. While our routers currently only support static routing algorithms such as dimensioned-ordered XY and table-based routing, the routing decision component in the router can easily be replaced with dynamic or custom routing logic. The routers have single-cycle latency due to input buffering. We can also register the router's output with an elastic buffer, trading off latency against an easier timing closure of very long routing channels. Finally, the internal switch of the router was optimized to disable loopbacks and impossible connections in XY-Routing.

\section{Compute Tile}
\begin{figure}[htbp]
    \centering
    \includegraphics[width=\columnwidth]{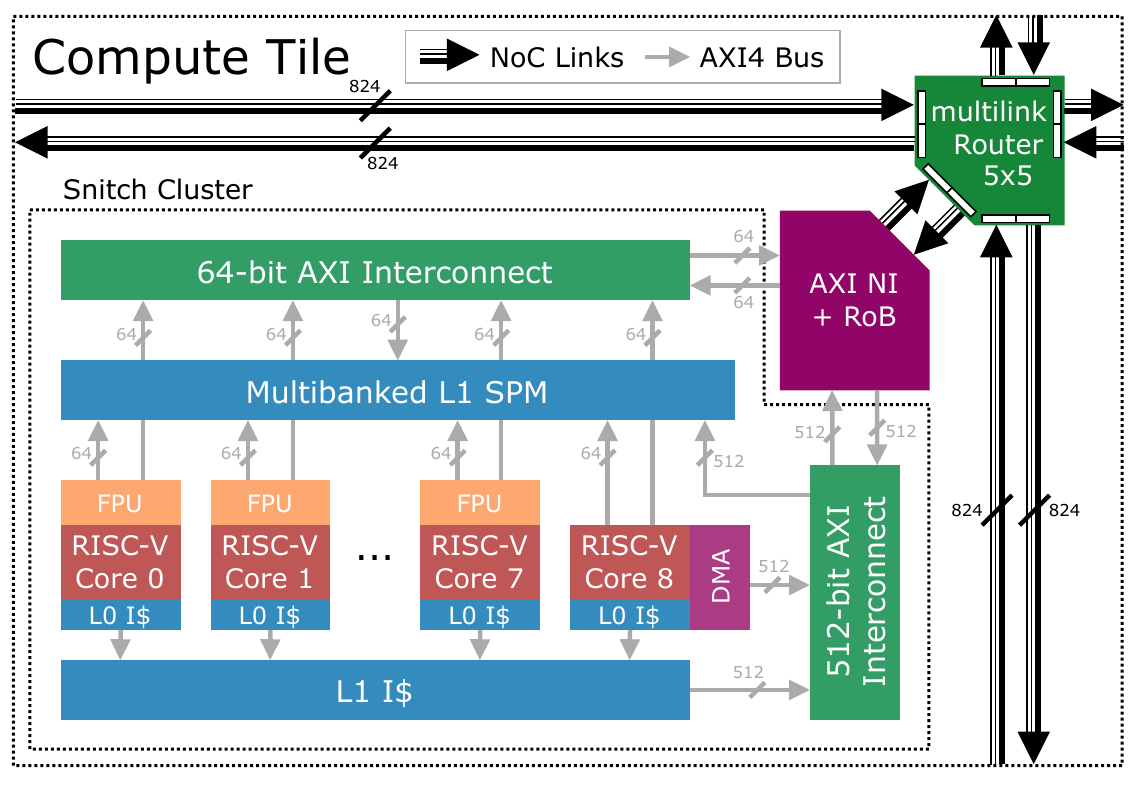}
    \caption{Compute tile consisting of a Snitch cluster with 9 RISC-V cores, an \gls{AXI}-\gls{NI} attached to a wide \SI{512}{\bit} and narrow \SI{64}{\bit} \gls{AXI} bus, and separate 5$\times$5 router for each physical link in the narrow-wide network.}
    \label{fig:cluster}
\end{figure}

As a case study of our \gls{NOC}, we integrated our \gls{NI} and router into a Snitch cluster~\cite{snitch}, forming a compute tile as shown in \Cref{fig:cluster}.
We configured the router to use XY-Routing with 5$\times$5 ports, one local port to the cluster, and one to each cardinal direction. The compute tile can be replicated to scale up to large on-chip multi-cluster systems like \textit{Manticore}~\cite{manticore} with an integrated \gls{NOC}. The compute cluster has 8 RISC-V cores with coupled \glspl{FPU} and a \gls{DMA} controlled by an additional RISC-V core. The \gls{SPM} and instruction cache are shared amongst all cores and are \SI{128}{\kilo\byte} and \SI{8}{\kilo\byte} large, respectively. The cluster-internal interconnect shown in green in \Cref{fig:cluster} consists of a wide 512-bit \gls{AXI} bus used by the \gls{DMA} and L1 instruction cache to fetch large chunks of data. Second, the RISC-V cores use a narrow 64-bit \gls{AXI} bus for single-word remote accesses. Both narrow and wide \gls{AXI} buses have initiator and target ports, meaning the cluster-internal \gls{SPM} is also accessible remotely by other clusters. We attached our \gls{AXI}-\gls{NI} to the narrow and wide \gls{AXI} buses with the mapping described in \Cref{tab:channels}. We configured \glspl{ROB} of size \SI{8}{\kilo\byte}\footnote{The size of the wide \gls{ROB} was chosen to allow at least 2 outstanding bursts of maximum size (\SI{4}{\kilo\byte}) for a sustained data flow.} for the wide and \SI{2}{\kilo\byte} for  the narrow \gls{AXI} bus.

\begin{figure*}[!htbp]
    \centering
    \includegraphics[width=0.9\textwidth]{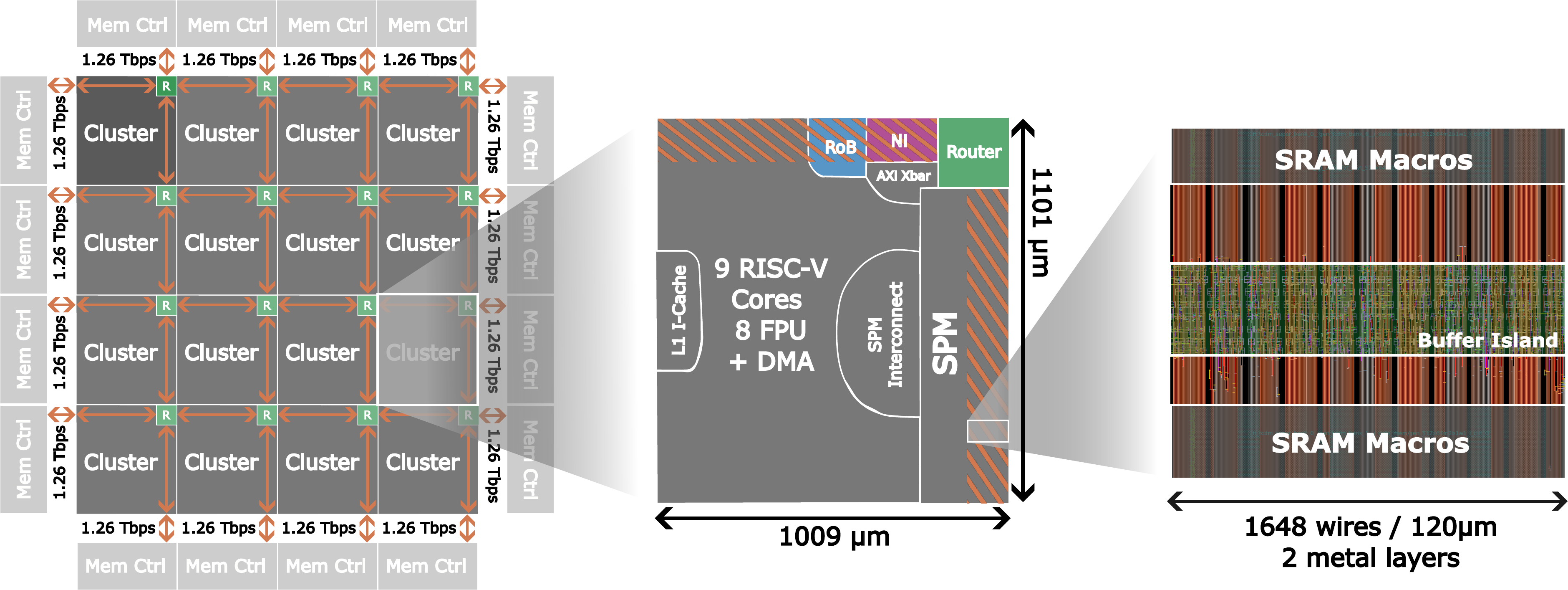}
    \begin{minipage}[t]{0.04\textwidth}
    
    \end{minipage}\hfill
    \begin{minipage}[t]{0.30\textwidth}
        \subcaption{\textbf{Compute Mesh:} Illustration of a system with a 4$\times$4 \gls{NOC} mesh embedded in abutted compute tiles with memory controllers at the boundary.}
        \label{fig:pnr_compute_mesh}
    \end{minipage}\hfill
    \begin{minipage}[t]{0.32\textwidth}
        \subcaption{\textbf{Compute Tile:} Annotated placed and routed design of a single compute tile. The physical routing channels over the tile components are annotated in orange.}
        \label{fig:pnr_compute_tile}
    \end{minipage}\hfill
    \begin{minipage}[t]{0.28\textwidth}
        \subcaption{\textbf{Physical Routing Channels:} Buffer islands between SRAM macros for refueling.}
        \label{fig:pnr_routing}
    \end{minipage}\hfill
    \begin{minipage}[t]{0.04\textwidth}

    \end{minipage}
    \caption{Annotated physical implementation of a \gls{NOC} connecting a mesh of compute tiles in \textsc{GlobalFoundries'} \SI{12}{\nano\meter}.}
    \label{fig:pnr}
\end{figure*}

\section{Physical Implementation}
We used Synopsys Fusion Compiler 2022.03 to synthesize, place, and route the compute tile and the embedded \gls{NOC} router and links in \textsc{GlobalFoundries'} \SI{12}{\nano\meter} FinFet technology. \Cref{fig:pnr_compute_tile} shows the annotated tile floorplan. 
The timing closes at a frequency of \SI{1.23}{\giga\hertz} in typical conditions (TT, \SI{0.8}{\volt}, \SI{25}{\celsius}), corresponding to a delay of 70 \glspl{FO4}. The router was synthesized, placed, and routed as a reusable black-box macro, and was instantiated in the upper right corner of the compute tile.
We reserved four upper metal layers to accommodate the horizontal and vertical physical routing channels, which are routed along the top and right edge of the compute tile.
By placing the \glspl{SRAM} of the \gls{SPM} and \gls{ROB}, which only obstruct lower metal layers, underneath the horizontal and vertical routing channels, we effectively minimized the overhead of the physical routing channels.
As detailed in \Cref{tab:channels}, a duplex channel requires approximately 1600 wires to be routed.
Assuming a near 100\% routing track utilization with some margin for the power grid and using two of the four metal layers with preferred routing direction, the routing channel occupies a slice of \SI{120}{\micro\meter}.
 
After floorplanning and power grid insertion, we routed the horizontal and vertical routing channels on the four reserved layers.
To bridge the large distances of the routing channels, we reserved space between the \gls{SRAM} macros for inserting buffer islands at regular distances, where the wires can \textit{refuel}\footnote{the term \textit{refueling} is used here to describe buffering very long signals routed on the upper layers to improve timing. Those signals need to go down to the logic layer to connect to the buffer cells} as shown in \Cref{fig:pnr_routing}.
We used a two-cycle router with output buffers to increase the timing budget of the routing channels.
For the given hard macro with \SI{1}{\milli\meter} long sides, timing analysis has shown that three sets of buffers are sufficient for the worst corner to satisfy transition time limitations.
To close timing at the top level, the tile's input and output timing has been constrained to allow multiple such hard macros to be abutted. This way, tiles can be arranged in a mesh to scale up without any frequency degradation, as illustrated in \Cref{fig:pnr_compute_mesh}. Memory controllers can be placed on the mesh boundary and connected to the \gls{NOC}.

\section{Results}
We evaluate the proposed design for power, performance, and area. The performance measurements were done using cycle-accurate simulations in \textsc{Siemens/Mentor QuestaSim 2022.3}. The area results were extracted from the post-layout netlist, and the power measurements were performed with \textsc{Synopsys PrimeTime 2022.3} using a post-layout simulation in typical conditions (TT, \SI{0.8}{\volt}, \SI{25}{\celsius}) at a frequency of \SI{1.23}{\giga\hertz}.

\begin{figure*}
    \centering
    \begin{minipage}[b]{0.48\textwidth}
        \centering
        \begin{tikzpicture}
        \begin{axis}[
            xmin=0,
            xmax=5,
            ymin=0,
            ymax=100,
            width=\textwidth,
            height=5cm,
            xlabel={\#Wide Interference Burst Transactions},
            ylabel={Narrow Latency [cycles]},
            legend columns=1,
            legend style={at={(axis cs: 0.2, 40)}, anchor=south west},
            xticklabels={0, 2, 4, 8, 16, 32, 64},xtick={0,...,6},
            xticklabel style = {font=\footnotesize},
            yticklabel style = {font=\footnotesize},
            y label style={at={(axis description cs:-0.07,.5)},rotate=0,anchor=south}
        ]
            \addplot [mark=square, color=PULPOrange, very thick] table [x expr=\coordindex, y=narrow_read_lat, col sep=comma] {measurements/lat_one_dir_wo.csv};
            \addlegendentry{wide-only}
            \addplot [mark=asterisk, color=PULPOrange, very thick] table [x expr=\coordindex, y=narrow_read_lat, col sep=comma] {measurements/lat_two_dir_wo.csv};
            \addlegendentry{wide-only (bidir)}
            \addplot [mark=square, color=PULPBlue, very thick] table [x expr=\coordindex, y=narrow_read_lat, col sep=comma] {measurements/lat_one_dir_nw.csv};
            \addlegendentry{narrow-wide}
            \addplot [mark=asterisk, color=PULPBlue, very thick] table [x expr=\coordindex, y=narrow_read_lat, col sep=comma] {measurements/lat_two_dir_nw.csv};
            \addlegendentry{narrow-wide (bidir)}
        \end{axis}
        \end{tikzpicture}
        \subcaption{Latency of narrow transactions with increasing interference of wide burst traffic. \textsc{NumNarrowTrans}=100}
        \label{fig:latency}
    \end{minipage} \hfill
    \begin{minipage}[b]{0.48\textwidth}
        \centering
        \begin{tikzpicture}
        \begin{axis}[
            name=plot1,
            xmin=0,
            xmax=6,
            ymin=50,
            ymax=90,
            width=\textwidth,
            height=5cm,
            xlabel={\#Narrow Interference Transactions},
            ylabel={Effective Wide BW [\%]},
            legend style={at={(axis cs: 0.2, 52)}, anchor=south west},
            legend columns=1,
            xticklabels={0, 2, 4, 8, 16, 32, 64},xtick={0,...,6},
            xticklabel style = {font=\footnotesize},
            yticklabel style = {font=\footnotesize},
            y label style={at={(axis description cs:-0.06,.5)},rotate=0,anchor=south}
        ]
            \addplot [mark=square, color=PULPBlue, very thick] table [x expr=\coordindex, y=wide_read_bw, col sep=comma] {measurements/bw_one_dir_nw.csv};
            \addlegendentry{narrow-wide}
            \addplot [mark=asterisk, color=PULPBlue, very thick] table [x expr=\coordindex, y=wide_read_bw, col sep=comma] {measurements/bw_two_dir_nw.csv};
            \addlegendentry{narrow-wide (bidir)}
            \addplot [mark=square, color=PULPOrange, very thick] table [x expr=\coordindex, y=wide_read_bw, col sep=comma] {measurements/bw_one_dir_wo.csv};
            \addlegendentry{wide-only}
            \addplot [mark=asterisk, color=PULPOrange, very thick] table [x expr=\coordindex, y=wide_read_bw, col sep=comma] {measurements/bw_two_dir_wo.csv};
            \addlegendentry{wide-only (bidir)}
        \end{axis}
        \end{tikzpicture}
        \subcaption{Effective bandwidth utilization of wide burst transactions with increasing interference of narrow transaction. \textsc{NumWideTrans}=16}
        \label{fig:bw}
    \end{minipage} \hfill
    \caption{Latency \& Bandwidth  comparison of cluster-to-cluster accesses between different \gls{NOC} configurations (narrow-wide and wide-only links) for unidirectional and bidirectional (bidir) traffic. \textsc{BurstLen}=16}
    \label{fig:latency_bw}
\end{figure*}
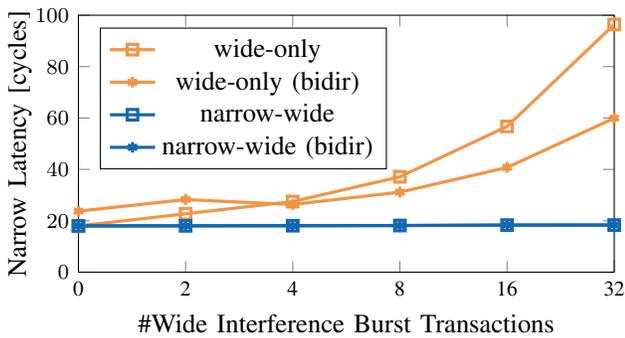
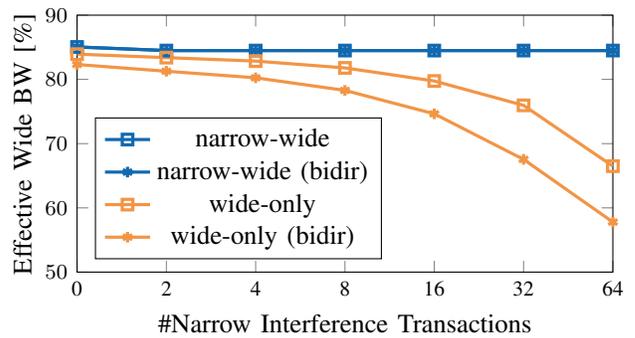

\subsection{Latency}
We measured the zero-load access latency of a compute tile to an adjacent tile. 
The combined latency of sending a request and receiving a response is 18 cycles.
During a tile-to-tile round-trip, a router is traversed four times, yielding a latency of eight cycles, while the NI contributes an additional cycle of latency.
The remaining nine cycles can be attributed to cluster-internal cuts and memory access latency.

Apart from the zero-load latency, we also measured the latency of latency-sensitive narrow transactions during bandwidth injection from the wide \gls{AXI}, which is a prevalent scenario in heterogeneous systems. As is shown in \Cref{fig:latency}, increasing congestion on a link can result in severe latency degradation of up to $5\times$, in a network with a single wide-only link. The burst transfers of the wide network cause very high link utilization, effectively starving the narrow latency-sensitive transactions. On the other hand, our approach of separating heterogeneous traffic onto multiple physical links proves to be very robust to heterogeneous traffic, with virtually no latency degradation on the narrow network.

\subsection{Bandwidth}
Wide burst-based \gls{DMA} transfers are more latency tolerant due to the ability to issue multiple outstanding transactions. However, they require a sustained high bandwidth data flow. Our \gls{NOC}'s wide link achieves a peak bandwidth of \SI{629}{\giga\bitpersecond} (\SI{1.26}{\tera\bitpersecond} duplex), operating at a relatively modest frequency of \SI{1.23}{\giga\hertz}. Arranged in a mesh, as depicted in \Cref{fig:pnr_compute_mesh}, our \gls{NOC} can deliver massive bandwidth at the boundary for traffic directed toward memory controllers and I/O. For instance, the aggregate bandwidth at the boundary of a 7$\times$7 mesh amounts to \SI{4.4}{\tera\byte\per\second}, exceeding the available memory and I/O bandwidth of state-of-the-art GPUs such as NVIDIA H100\cite{nvidia_hopper_h100}. 

Clearly, the effective bandwidth can be much lower than the peak bandwidth if the utilization of the link is low, such as when small messages like \gls{AXI} AR/AW requests and B responses are sent over the same wide link used by the read and write messages. These small messages only use a fraction of the wide link, consequently reducing the effective bandwidth. The use of narrow-wide links addresses this issue by sending small messages over separate narrow links. We measured the impact of our approach on the effective bandwidth of the wide link by comparing it to an implementation with a single wide-only link. We varied the amount of narrow traffic injected into the network and present the results in \Cref{fig:bw}. Our wide link achieves an effective bandwidth utilization of \SI{85}{\percent} and is very robust to bandwidth degradation compared to a single wide-only link.

\subsection{Area}
\begin{figure}
    \centering  
    \begin{subfigure}[b]{\columnwidth}
    \centering
    \includegraphics[width=\textwidth]{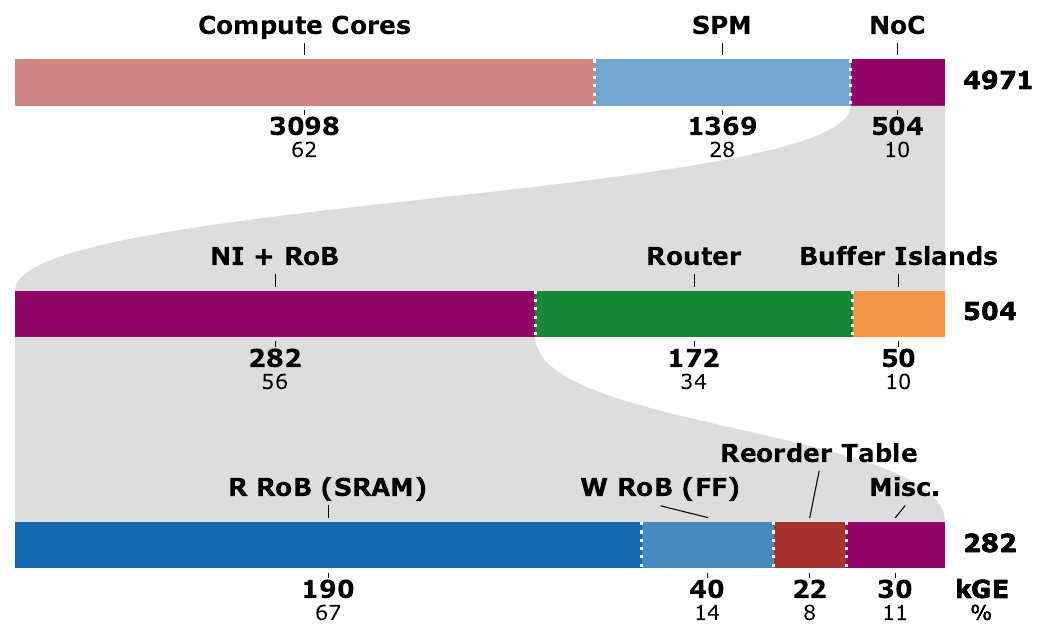}
    \subcaption{Area breakdown}
    \label{fig:area_breakdown}
  \end{subfigure}\hfill
  \begin{subfigure}[b]{\columnwidth}
    \centering
    \includegraphics[width=\textwidth]{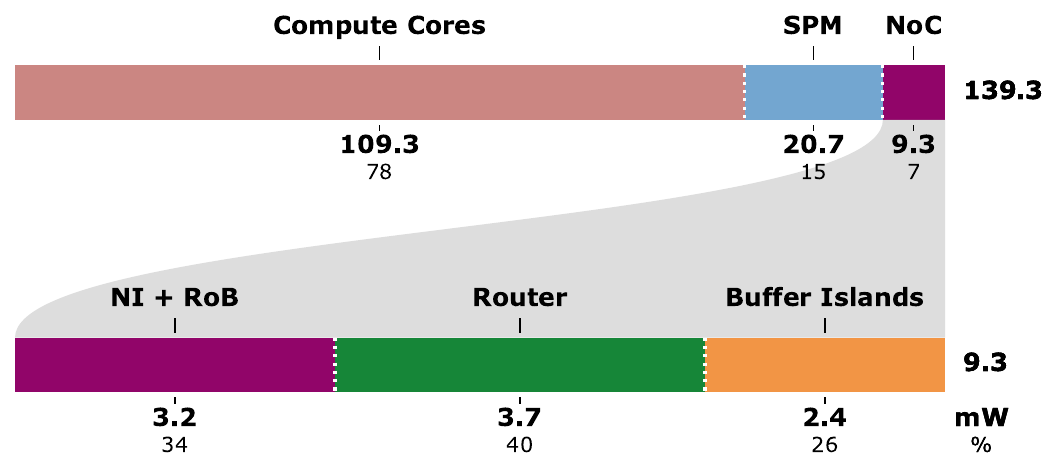}
    \subcaption{Power breakdown}
    \label{fig:power_breakdown}
  \end{subfigure}\hfill
  \caption{Area \& Power results: (a) Area breakdown of the compute tile with an integrated router and \gls{NI} (b) Power breakdown of the compute tile and \gls{NOC} during a \SI{1}{\kilo\byte} transfer.}
\end{figure}

\Cref{fig:area_breakdown} shows an area breakdown of the compute tile, which measures approximately \SI{5}{\mega\gate} in size. The \gls{NOC} components, comprising the router, \gls{NI}, \gls{ROB}, and buffer islands, have a complexity of \SI{500}{\kilo\gate}, accounting for a mere \SI{10}{\percent} of the total compute tile area. Effectively, the \gls{NOC} with the routing channels covers roughly a quarter of the entire floorplan. However, as it is routed on the upper layers above the \gls{SRAM} macros, its impact on the compute tile area remains minimal. The \gls{NOC}'s size is primarily governed by the \gls{NI} and its \glspl{ROB}, which are essential for ensuring a sustained high-bandwidth dataflow while complying with the \gls{AXI} ordering requirements. The \SI{8}{\kilo\byte} and \SI{2}{\kilo\byte} \glspl{ROB} for the wide and narrow read responses were implemented as \gls{SRAM}, whereas the Reorder Table and write responses use \gls{SCM}, as \gls{AXI} write responses are too small to justify the use of \gls{SRAM}.

\subsection{Energy Efficiency}
We conducted post-layout power simulations to measure the energy efficiency of the \gls{NOC} by issuing a single \gls{DMA} transfer of size \SI{1}{\kilo\byte} and extracting the power of each component, as illustrated in \Cref{fig:power_breakdown}. During the transfer, all RISC-V cores, except for the core programming the \gls{DMA}, remain idle. The total power consumption of the tile is \SI{139}{\milli\watt}, of which only \SI{7}{\percent} is consumed by the \gls{NOC}. This indicates that even with an idle compute cluster, the power consumption of the \gls{NOC} is almost negligible. As a result, the limiting factor for scaling up to multi-cluster systems will not reside in the \gls{NOC} but in the compute logic. Additionally, we computed the energy consumed by the router and routing buffers for moving \SI{1}{\kilo\byte} across the tile, which amounts to \SI{198}{\pico\joule}, resulting in an energy efficiency of \SI{0.19}{\pico\joule\per\byte\per\hop}.

\section{Related Works}
\gls{NOC} design has been an active area of research for more than two decades. A comparative overview of recent \glspl{NOC} is provided in \Cref{tab:related_works}. However, only a limited number of studies have explored the design of extremely wide physical links for high-bandwidth communication. Ruche channels~\cite{ou_implementing_2020} were introduced to enhance latency and bandwidth by incorporating additional links that bypass routers. Nevertheless, spanning large distances without violating timing paths at the expense of latency poses a challenge for larger tile sizes, and router complexity increases with higher radix routers. Moreover, the doubled bisection bandwidth does not lead to a $2\times$ memory bandwidth improvement, as memory controllers are typically situated at the periphery, which implies that traffic will be predominantly routed through the ruche channels.

\glspl{NOC} addressing diverse traffic patterns have also been investigated~\cite{das_design_2009} by employing wide channels to transmit multiple small flits concurrently, thereby enhancing the bandwidth and latency of various traffic patterns. However, this approach incorporates arbitration logic in the routers and fails to alleviate traffic congestion on the wide link, unlike our solution with distinct narrow and wide physical links.

\glspl{NOC} with \gls{AXI} support have been proposed using \gls{AXI} matrices~\cite{kurth_open-source_2022}. However, to maintain \gls{AXI} compliance, each matrix must monitor outstanding requests. ID remappers need to be implemented to counteract the increasing ID width after each hop, ultimately compromising the scalability of such systems. \gls{AXI}-\gls{NI} proposals~\cite{axi_reorder, ebrahimi_high-performance_2010} that adhere to the reordering requirements exist but as closed-source solutions and lack quantified and detailed information on the physical implementation of the \gls{NOC}.

\newcommand*\rot[1]{\hbox to1em{\hss\rotatebox[origin=br]{-60}{#1}}}
\newcommand*\haeggli{\Circled[inner color=white, outer color=white, fill color=PULPGreen]{\cmark}}
\newcommand*\chruezli{\Circled[inner color=white, outer color=white, fill color=PULPRed]{\xmark}}
\newcommand*\welleli{\Circled[inner color=white, outer color=white, fill color=PULPOrange, inner ysep=6pt]{$\approx$}}

\begin{table}[]
    \centering
    \caption{Comparison of FlooNoC with State-of-the-Art \glspl{NOC}}
    \begin{tabular}{@{}l c c c l l l l@{}}
    \toprule
      \textbf{Work} & \textbf{\rot{Link DW [bit]}} & \textbf{\rot{Frequency \SI{}{[\giga\hertz]}}} & \textbf{\rot{Link BW [\SI{}{\giga\bitpersecond}]}} & \textbf{\rot{Open Source}} & \textbf{\rot{Fully AXI comp.}} & \textbf{\rot{Mult. outst. trans.}} & \textbf{\rot{Physical Impl.}}\\
    \midrule
    FlexNoC~\cite{flexnoc} & n.A.$^\dag$ & n.A.$^\dag$ & n.A.$^\dag$ & \chruezli & \haeggli & \haeggli & \welleli$^\dag$ \\
    CoreLink~\cite{arm_noc} & $\leq$ 512 & 1 & 512 & \chruezli & \haeggli & \haeggli & \welleli$^\dag$ \\
    ESP~\cite{esp_noc_soc} & 5$\times$64 & 0.8 & 281 & \haeggli & \chruezli & \chruezli & \haeggli \\
    Constellation~\cite{zhao_constellation_2022} & 64 & 0.5 & 32 & \haeggli & \welleli$^a$ & \welleli$^b$ & \haeggli \\
    OpenPiton~\cite{openpiton} & 3$\times$64 & 1 & 192 & \haeggli & \welleli$^c$ & \haeggli & \haeggli \\
    Celerity~\cite{celerity} & 80 & 1 & 80 & \haeggli & \chruezli & \chruezli & \haeggli \\
    \gls{AXI}-XP~\cite{kurth_open-source_2022} & \textbf{512/64} & 1 & 512 & \haeggli & \haeggli & \haeggli & \welleli$^d$ \\
    \textbf{This work} & \textbf{512/64} & \textbf{1.23} & \textbf{629} & \haeggli & \haeggli & \haeggli & \haeggli \\
    \bottomrule
    \end{tabular}
    \label{tab:related_works}
    \footnotesize{$^\dag$Has not been benchmarked in open literature, $^a$No \gls{AXI} reordering},\\
    $^b$only 1 transaction per ID, $^c$Only \gls{AXI}-Lite, $^d$not scalable
\end{table}

\section{Conclusion}
In this paper, we have presented the design of a novel \gls{NOC} characterized by wide parallel physical channels, which do not require frequency multiplication to sustain the bandwidth of wide \gls{AXI} channels. Our \gls{NOC} achieves a bandwidth of \SI{629}{\giga\bitpersecond} per link at a modest frequency of \SI{1.23}{\giga\hertz} and a delay of 70 \glspl{FO4}. To the best of our understanding, our \gls{NOC} is the first fully-open source design to offer an end-to-end solution that is fully compliant with the \gls{AXI} standard, incorporating a \gls{NI} that satisfies its reordering prerequisites. We have proposed the use of separate narrow and wide links to cater to the demands of both high-bandwidth and low-latency traffic within the network. To demonstrate our \gls{NOC} in a realistic high-bandwidth \gls{SOC} design, we embedded it in a 9-core RISC-V cluster with a wide, high-bandwidth 512-bit \gls{DMA} channel and a 64-bit latency critical port for remote load/stores by the cores. This case study demonstrates the practicality of wide physical channels: The \gls{NOC} occupies a mere \SI{10}{\percent} of the compute tile. Contributing only \SI{7}{\percent} to the power consumption of a computational tile and achieving an energy-efficiency of \SI{0.19}{\pico\joule\per\byte\per\hop}, our \gls{NOC} design effectively demonstrates that it will not become the bottleneck in large multi-cluster systems.

As part of our future work, we aim to perform a more detailed comparison between our proposal and designs based on virtual channels, which would offer further insight into the benefits of our approach.

\ifblind
\else
\section*{Acknowledgments}

This work has been supported in part by ‘The European Pilot’ project under grant agreement No 101034126 that receives funding from EuroHPC-JU as part of the EU Horizon 2020 research and innovation programme.

\fi

\bibliographystyle{IEEEtran}
\bibliography{ieeetran.bib, references.bib, bibliography.bib}

\ifblind
\else
\begin{IEEEbiography}[{\includegraphics[width=1.05in,height=1.25in,clip,keepaspectratio]{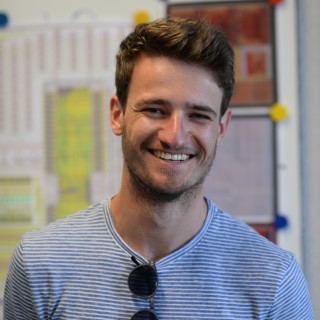}}]{Tim Fischer}
received his BSc and MSc in ``Electrical Engineering and Information Technology'' from the Swiss Federal Institute of Technology Zurich (ETHZ), Switzerland, in 2018 and 2021, respectively. He is currently pursuing a Ph.D. degree at ETH Zurich in the Digital Circuits and Systems group led by Prof. Luca Benini. His research interests include scalable and energy-efficient interconnects for both on-chip and off-chip communication.
\end{IEEEbiography}

\begin{IEEEbiography}[{\includegraphics[width=1.05in,height=1.25in,clip,keepaspectratio]{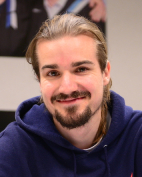}}]{Michael Rogenmoser}
received his BSc and MSc in ``Electrical Engineering and Information Technology'' from the Swiss Federal Institute of Technology Zurich (ETHZ), Switzerland, in 2020 and 2021, respectively. He is currently pursuing a Ph.D. degree in the Digital Circuits and Systems group of Prof. Benini. His research interests include fault-tolerant processing architectures and multicore heterogeneous SoCs for space.
\end{IEEEbiography}

\begin{IEEEbiography}[{\includegraphics[width=1.05in,height=1.25in,clip,keepaspectratio]{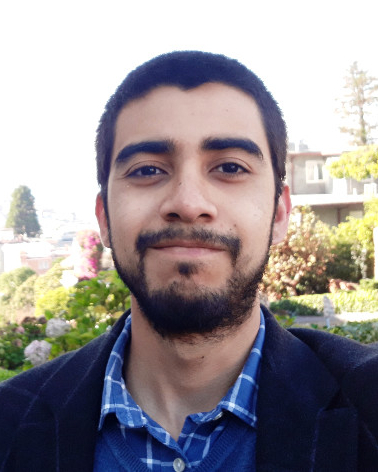}}]{Matheus Cavalcante}
received his M.Sc.\ degree in Integrated Electronic Systems from the Grenoble Institute of Technology (Phelma), France, in 2018. In the same year, he joined the Digital Circuits and Systems Group of ETH Zurich, Switzerland, where he is working toward a Ph.D.\ degree under the supervision of Prof.\ Luca Benini. Matheus' research interests include vector processing, high-performance computer architectures, and emerging VLSI technologies.
\end{IEEEbiography}

\begin{IEEEbiography}[{\includegraphics[width=1.05in,height=1.25in,clip,keepaspectratio]{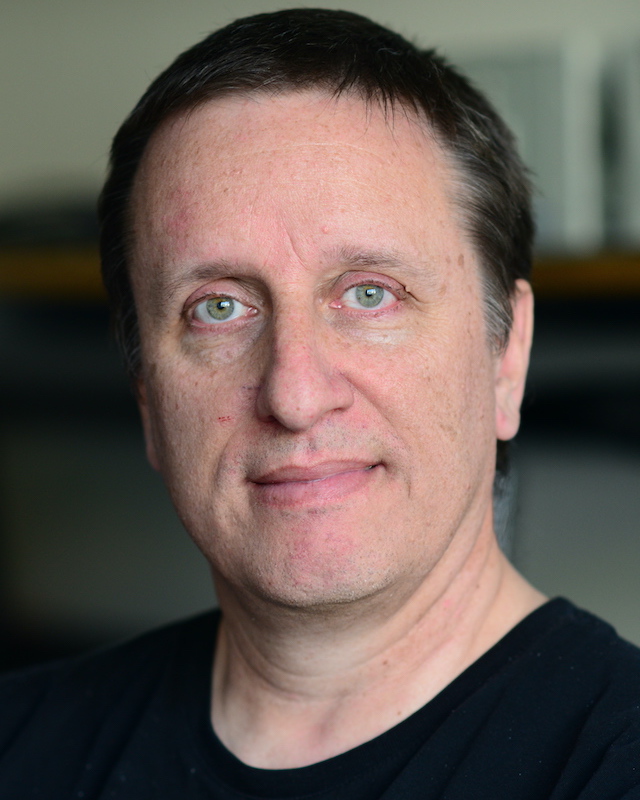}}]{Frank Gürkaynak}
received the BSc and MSc degrees in electrical engineering from the Istanbul Technical University, and the PhD degree in electrical engineering from ETH Zurich in 2006. He is currently working as a senior scientist with the Integrated Systems Laboratory, ETH Zurich. His research interests include digital low-power design and cryptographic hardware.
\end{IEEEbiography}

\begin{IEEEbiography}[{\includegraphics[width=1.05in,height=1.25in,clip,keepaspectratio]{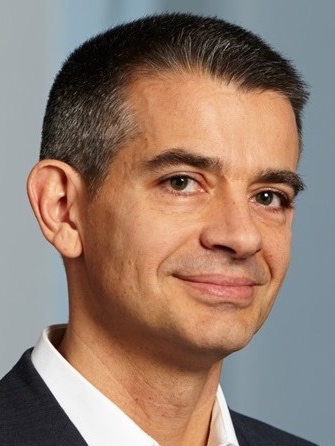}}]{Luca Benini}
holds the chair of digital Circuits and systems at ETHZ and is Full Professor at the Università di Bologna. He received a Ph.D. from Stanford University. Dr. Benini's research interests are in energy-efficient parallel computing systems, smart sensing micro-systems and machine learning hardware. He is a Fellow of the IEEE, of the ACM and a member of the Academia Europaea.
\end{IEEEbiography}
\fi

\end{document}